\documentclass[prd, onecolumn, showkeys, amsmath,amssymb,superscriptaddress,nofootinbib,11pt]{revtex4-2}
\usepackage{url}

\usepackage{epsfig}
\usepackage{amsfonts}
\usepackage{graphicx}
\usepackage{subfigure}
\usepackage{epsfig}
\usepackage{eepic}
\usepackage{amsmath}
\usepackage{amssymb}
\usepackage{color}
\usepackage{bbm}
\usepackage{dcolumn}% Align table columns on decimal point
\usepackage{bm}% bold math
\usepackage[normalem]{ulem}
\usepackage{mathrsfs}
\usepackage{bbold}
\usepackage{datetime}

\usepackage{overpic} % add text over picture
\usepackage{rotating}
\usepackage[usenames,dvipsnames]{xcolor}
\usepackage[colorlinks=true,citecolor=Magenta,linkcolor=Green,urlcolor=Green]{hyperref}
\usepackage{lipsum} % for mock text
\usepackage{tikz,tikz-3dplot}
\usetikzlibrary{shapes.geometric}
\usepackage{etoolbox} % for \appto
\usepackage[capitalize]{cleveref}
\usepackage{extarrows}
\usepackage{ragged2e}

\usepackage{soul}

\def\bc{\begin{center}}

\def\ec{\end{center}}
\def\be{\begin{eqnarray}}
\def\ee{\end{eqnarray}}
\definecolor{dyellow}{rgb}{1.,0.8,.0}
\definecolor{myblue}{rgb}{.1,.1,.7}
\definecolor{dcyan}{rgb}{.0,.6,.6}
\definecolor{dmagenta}{rgb}{0.6,0.0,0.6}
\definecolor{brown}{rgb}{0.6,0.2,0.}
\definecolor{darkblue}{rgb}{.0,.0,0.5}
\definecolor{darkred}{rgb}{0.75,0.0,0.0}
\definecolor{orange}{rgb}{1.,.6,.0}
\definecolor{dorange}{rgb}{0.8,.4,.0}
\definecolor{darkgreen}{rgb}{0.0,0.6,0.0}
\definecolor{purple}{rgb}{.4,.0,.4}
\definecolor{lightgrey}{rgb}{0.7, 0.7, 0.7}
\definecolor{grey}{rgb}{0.4, 0.4, 0.4}
\usepackage{geometry}
\usepackage[font=small,labelfont=bf,justification=raggedright]{caption}

\newcommand{\xdownarrow}[1]{%
  {\left\downarrow\vbox to #1{}\right.\kern-\nulldelimiterspace}
}
\newcommand{\xuparrow}[1]{%
  {\left\uparrow\vbox to #1{}\right.\kern-\nulldelimiterspace}
}

\definecolor{myred}{RGB}{189, 38, 49}

\begin{document}
\title{Topologically Protected Metastable States in Classical Dynamics}
\author{Han-Qing Shi} \email{by2030104@buaa.edu.cn}
\affiliation{Center for Gravitational Physics, Department of Space Science, Beihang University, Beijing 100191, China}
\author{Tian-Chi Ma} \email{tianchima@buaa.edu.cn (Corresponding author)}
\affiliation{Center for Gravitational Physics, Department of Space Science, Beihang University, Beijing 100191, China}
\author{Hai-Qing Zhang} \email{hqzhang@buaa.edu.cn (Corresponding author)}
\affiliation{Center for Gravitational Physics, Department of Space Science, Beihang University, Beijing 100191, China}
\affiliation{Peng Huanwu Collaborative Center for Research and Education, Beihang University, Beijing 100191, China}

\begin{abstract}
We propose that domain walls formed in a classical Ginzburg-Landau model can exhibit topologically stable but thermodynamically metastable states. This proposal relies on Allen-Cahn's assertion that the velocity of domain wall is proportional to the mean curvature at each point. From this assertion we speculate that domain wall behaves like a rubber band that can winds the background geometry in a nontrivial way and can exist permanently. We numerically verify our proposal in two and three spatial dimensions by using various boundary conditions. It is found that there are possibilities to form topologically stable domain walls in the final equilibrium states. However, these states have higher free energies, thus are thermodynamically metastable. These metastable states that are protected by topology could potentially serve as storage media in the computer and information technology industry.
\end{abstract}


\maketitle

\section{Introduction}
Topologically protected states are frequently studied in quantum physics in recent years, such as the gapless boundary excitations in intrinsic topological order due to the long-range entanglement \cite{fefferman2017topologically,zeng2019quantum}. They are robust against any local perturbations and are important to quantum computing \cite{nayak2008non,stern2013topological}. However, topologically protected states in classical physics are seldom explored. In this work we show that topologically protected states can also exist in classical dynamics, although they are metastable from the aspects of thermodynamics.

To make our ideas concrete, it is helpful to consider the simplest and most familiar system: Ginzburg-Landau model \cite{hohenberg2015introduction}.
Specifically, we adopt a time-dependent Ginzburg-Landau (TDGL) model to investigate the topologically protected metastable states in classical dynamics \cite{tang1995time,Kopnin2001}. TDGL model has a $Z_2$ symmetry of the real scalar field in the disordered phase.  Quenching it into an ordered phase, the previous $Z_2$ symmetry will spontaneously break and form the mosaic patterns which are the symmetry-breaking domains \cite{kibble1976topology,kibble1980some,zurek1985cosmological}. Inside each domain the scalar fields (order parameters) take the same sign.  
Therefore, domain walls turn out as the interfaces between different symmetry-breaking domains. As time evolves, domain walls will move, bend and merge according to the coarsening dynamics of TDGL \cite{bray2002theory}. TDGL model describes the dynamics of non-conserved order parameters, therefore, it belongs to the model A in the classification in \cite{hohenberg1977theory}.

Surface tension of the domain wall results in a tangent force along the interface. From Allen-Cahn \cite{allen1979microscopic}, the domain wall will move according to how the domain wall bends at that point. Specifically, the magnitude of the velocity is linearly proportional to the mean curvature, while the direction of the velocity is opposite to the normal direction. From Allen-Cahn's assertion, we further speculate that the movements of domain walls are similar to those of rubber bands in order that they can relax in an elastic way to the configurations with lowest free energies. Consequently, the domain wall will finally settle down with different topologies due to how the interface winds around the background geometry in the final equilibrium state. 

In the beginning of the evolution, we randomly fluctuate the system slightly and then quench the Hamiltonian to evolve the system. Consequently, the final configurations of domain walls will exist with probabilities. 
We find that in most cases the final states of the system have vanishing domain walls, corresponding to the ground state with lowest free energies. In this sense the system is stable thermodynamically. However, there are also possibilities that, in both periodic and Neumann boundary conditions, the domain walls will not vanish in the final equilibrium state. In two(three) dimensions with periodic boundary conditions, the background geometry is like a torus $T^2$($T^3$). Domain walls in this background have the chance to wind around the torus nontrivially and remain in this state permanently.  However, from the aspects of thermodynamics, this state is metastable since the free energy is not minimal. Interestingly, even with Neumann boundary conditions we can still have non-vanishing domain walls with nontrivial topologies in the final equilibrium state. This is similar to the case that a rubber band connects two moving points at opposite sides of a rectangle, while the points can only move along the two sides. Consequently, the rubber band will try to drag the moving points and finally arrange them in a straight line perpendicular to both sides (see the following).  Hence, we realize the topologically protected metastable states in the classical coarsening dynamics.

Since the domain wall separates the antiphase order parameters, i.e., order parameters with opposite signs, we can naturally expect that the topologically protected states may be used as storage media for computer and information technology industry. The advantage is that this kind of material is only of classical mechanics rather than quantum mechanics.

\section{Statement of the speculations}

%\subsection{Time-dependent Ginzburg-Landau model}
 For a non-conserved order parameter $\phi(\vec x,t)$, its equation of motion (EoM) can be described from the variance of the free energy as \cite{bray2002theory}
\be\label{eomnonoise}
\frac{\partial\phi(\vec x,t)}{\partial t}=-\Gamma\frac{\delta F[\phi]}{\delta\phi}
\ee
where $\Gamma$ is a positive kinetic coefficient and $F[\phi]$ is the free energy functional. Generically, for a TDGL model, $F[\phi]$ can be written  as
\be\label{freeenergy}
F[\phi]=\int d^dx\left(\frac12(\nabla\phi)^2+V(\phi)\right).
\ee
where the potential $V(\phi)$ is even under the permutation $\phi\to-\phi$ and has a structure like Mexican hat. The EoM can be readily obtained from Eqs.\eqref{eomnonoise} and \eqref{freeenergy}, 
\be\label{eqgl}
\frac{\partial\phi}{\partial t}+\Gamma\left(-\nabla^2\phi+\frac{\partial V}{\partial\phi}\right)=0. 
\ee
There exists a critical point which represents a phase transition from disordered phase to order phase. Therefore, the previous $Z_2$ symmetry in the disordered phase will spontaneously break in the ordered phase. Consequently, defects will turn out due to the spontaneous $Z_2$ symmetry breaking, such as domain walls (or interfaces) which separate the antiphase order parameters $\pm\phi$. (In one-dimensional space, the defects are kinks.) \cite{vachaspati2007kinks}

\begin{figure}[t]
\centering
\includegraphics[trim=0cm 0cm 0cm 0cm, clip=true, scale=0.4]{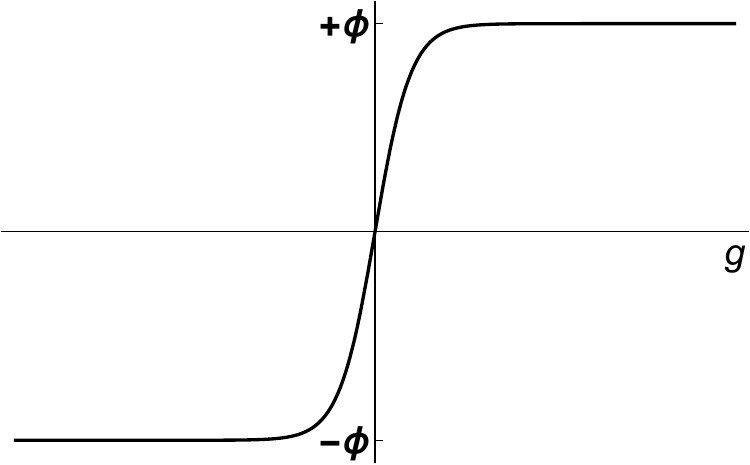}
\put(-150,80){(a)}
~~~~
\includegraphics[trim=4.5cm 1.5cm 6.cm 1.cm, clip=true, scale=0.55]{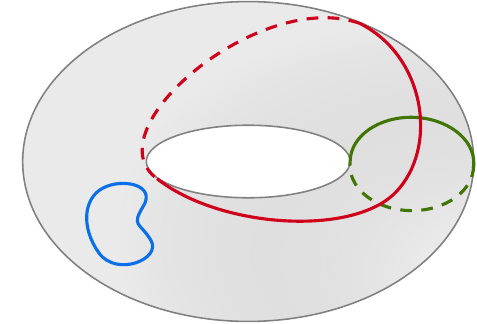}\vspace{0.4cm}
\put(-120,80){(b)}\\
\includegraphics[trim=0cm 2.cm 0.5cm 1.5cm, clip=true, scale=0.49]{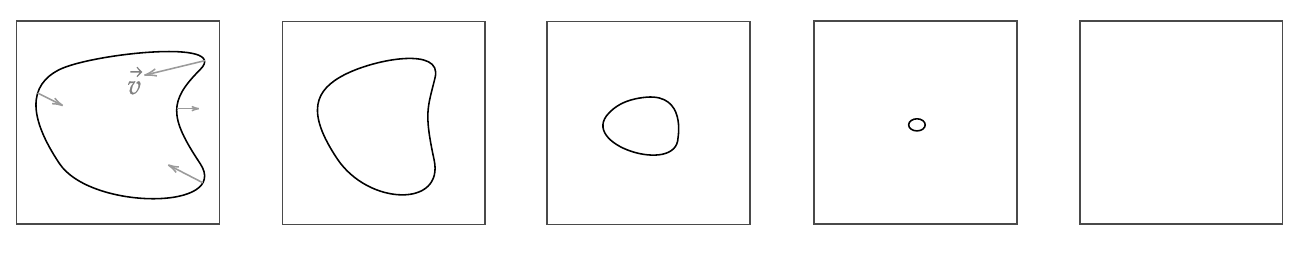}
\put(-302,60){(c)}
\caption{(a). Sketchy profile of the order parameter along the normal direction of the domain wall in the ordered phase. The slope is the cross section of the domain wall; (b). Three typical interfaces with distinct topologies on a torus. (c) Illustrative evolutions of a closed domain wall with trivial topology. Gray arrows represent the velocity vectors of the interfaces. This domain wall will gradually shrink and finally disappear during the evolution. For illustrative purpose, we simply use the same colors to represent the antiphase order parameters separated by the domain wall. }\label{fig1}
\end{figure}

During the coarsening dynamics, the order parameters along the normal direction $\vec n$ of the interface almost have identical shapes \cite{allen1979microscopic}. Fig.\ref{fig1}(a) shows the cross sectional profile of the antiphase order parameter along the normal direction of the interface. The profile behaves like $\phi\simeq\tanh(g)$ where $g$ is the coordinates along the normal direction \cite{weinberg1992classical}.   According to Allen-Cahn \cite{allen1979microscopic}, the velocity $\vec v$ of the interface is linearly proportional to the mean curvature $H$ of the domain wall, i.e., $|\vec v|\propto H$.  The direction of the velocity is opposite to the direction of the normal vector $\vec n$ at that point as indicated in Fig.\ref{fig1}(c).  

From Allen-Cahn's assertion that $|\vec v|\propto H$, we further speculate that {\it the domain wall in the TDGL model \eqref{eqgl} resembles a rubber band with zero natural length}.  
 ``Zero natural length'' implies that the domain wall can shrink and disappear if there is no obstacles. Therefore,  a closed and topologically trivial interface will gradually shrink and finally disappear.  For example, the blue loop in Fig.\ref{fig1}(b) is topologically trivial. We sketch its dynamics in Fig.\ref{fig1}(c), in which gray arrows indicate the velocity vectors at the interface. Arrow directions are the velocity directions opposite to the normal directions, while the arrow lengths imply the magnitudes of the velocities. Because of $|\vec v|\propto H$, the convex parts of the curve will move inwards while the concave parts will move outwards, behaving like an elastic rubber band with zero natural length. Consequently, after some time, an interface with uneven loop will gradually bend and shrink, then become more even and finally disappear. See the subsequent sketchy pictures in Fig.\ref{fig1}(c). 

On the other hand, vanishing of a trivial loop is consistent with the requirements of the lowest free energy of the system. In the ordered phase the domain wall is like the function $\tanh(g)$ along the normal direction (see Fig.\ref{fig1}(a)). Therefore, the gradient energy of the interface $(\nabla\phi)^2$ in the free energy functional Eq.\eqref{freeenergy} are almost identical at each point of the interface \cite{allen1979microscopic}. Hence the free energy of the system depends crucially on the lengths of the interface since other parts of the order parameter are uniform. Therefore, after a long enough time the system needs to relax in a state with lowest free energy, i.e., with vanishing domain walls. This is why topologically trivial interfaces will shrink and finally disappear from the aspects of thermodynamics. 

However, if the interface is topologically nontrivial, the above statement should be altered.  For example, in Fig.\ref{fig1}(b) the green and red loops are belonging to this type. They wind around the torus and cannot be untangled. Therefore, similar to our discussions above, we can conclude that the final profile of the interface should be like a rubber band surrounding the torus tightly. It cannot shrink or vanish due to the topological obstacles. However, from the analysis of the thermodynamics, this state does not have the lowest free energy since the interface length is nonzero. They are metastable states from the aspects of thermodynamics! Therefore, we can conclude that these metastable states are in fact stable due to their nontrivial topologies, namely they are topologically protected metastable states.

\section{Results}
\subsection{Quench the TDGL model}
 In order to verify our speculations above, we consider the following TDGL models with the potential $V=\phi^4/8-\epsilon(t)\phi^2/4$,
\be\label{GLeq2}
\dot\phi-\Gamma\left(\nabla^2\phi-\frac12\left(\phi^3-\epsilon(t)\phi\right)\right)=0. 
\ee
where $\epsilon(t)$ is a time-dependent parameter to control the shape of the potential. In the equilibrium state, as $\epsilon<0$ the potential has a single minimum at $\phi=0$, therefore, the solution with vanishing $\phi$ is preferred. However as $\epsilon>0$ the potential has two minima at $\phi=\pm\sqrt\epsilon$ respectively, and the preferred solution is $\phi=\pm\sqrt\epsilon$. Therefore, $\epsilon=0$ is the critical point of the phase transition from vanishing $\phi$ to finite $\phi$. Dynamically, if we quench the parameter $\epsilon(t)$ from negative to positive, phase transitions will occur from disordered phase to ordered phase. Therefore, the previous $Z_2$ symmetry of the real scalar field $\phi$ will spontaneously break into positive or negative values. Consequently, domain walls will form as the interfaces of the antiphase order parameters. For simplicity, we quench $\epsilon$ linearly as $\epsilon(t)={t}/{\tau_Q}$, where $\tau_Q$ is the quench rate and we have fixed it as $\tau_Q=e$. Specifically, we start the quench from $t/\tau_Q=-2$ to $t/\tau_Q=10$, then we stop the quench and maintain the parameter $\epsilon$ at $\epsilon_f=10$. Until the final equilibrium state, the order parameter will take the values $\phi=\pm\sqrt{\epsilon_f}=\pm\sqrt{10}$ except the interfaces connecting them.  To analytically solve the Eq.\eqref{GLeq2} is formidable, hence we take advantage of numerical simulations. In the numerics, we set $\Gamma=1/5$ and the length of the system as $L=100$.  In the time direction, the 4-th Runge-Kutta methods are adopted with the time step $\Delta t=0.02$. In the spatial direction we have respectively used periodic boundary conditions or Neumann boundary conditions (see the following).  In the initial state the random fluctuations are taken as the white noise $\xi(t,\vec x)$ satisfying $\langle \xi(t,\vec x)\rangle=0$ and $\langle \xi(t',\vec x')\xi(t,\vec x)\rangle=\eta\delta(\vec x'-\vec x)\delta(t'-t)$ with $\eta=10^{-6}$. \footnote{{We choose $\Gamma=1/5$ since as we have multiple trials, we find that $\Gamma$ cannot be too large. Other small values of $\Gamma$ will also work similarly. $\eta$ is the amplitude of the initial seeds which should be seen as small fluctuations. Other small values of $\eta$ will also work similarly. The system size $L=100$ which is much larger than the correlation length of the fields. The correlation length of the fields is roughly the size of the width of the domain wall. Therefore, we see that $L=100$ is suitable for the emergence of the domain walls. For other parameters in the numerical simulations, we consider both the stability and efficiency of the codes. }}

\subsection{Domain walls in two dimensions: I. Periodic Boundary Conditions}
%{\it Domain walls in two dimensions: I. Periodic Boundary Conditions} -- 
We first investigate the formation of domain walls in two spatial dimensions with periodic boundary conditions, thus the topology of the background geometry is like a two-torus $T^2$ as indicated in Fig.\ref{fig1}(b). Numerically, we Fourier decompose the two spatial lengths to $401$ grids.  We use the symbol $(w_x,w_y)$ to represent the windings that the domain wall may surround the periodic $x$ and $y$ directions. Since we are working with periodic boundary conditions, the domain walls always appear in pairs \cite{del2018universal,PhysRevLett.124.240602,delCampo:2021rak}. With periodic boundary conditions, $(w_x,w_y)$ just counts windings of half of the pairs. 

\begin{figure}[t]
\centering
\includegraphics[trim=0.2cm 4.1cm 0cm 5.1cm, clip=true, scale=0.5]{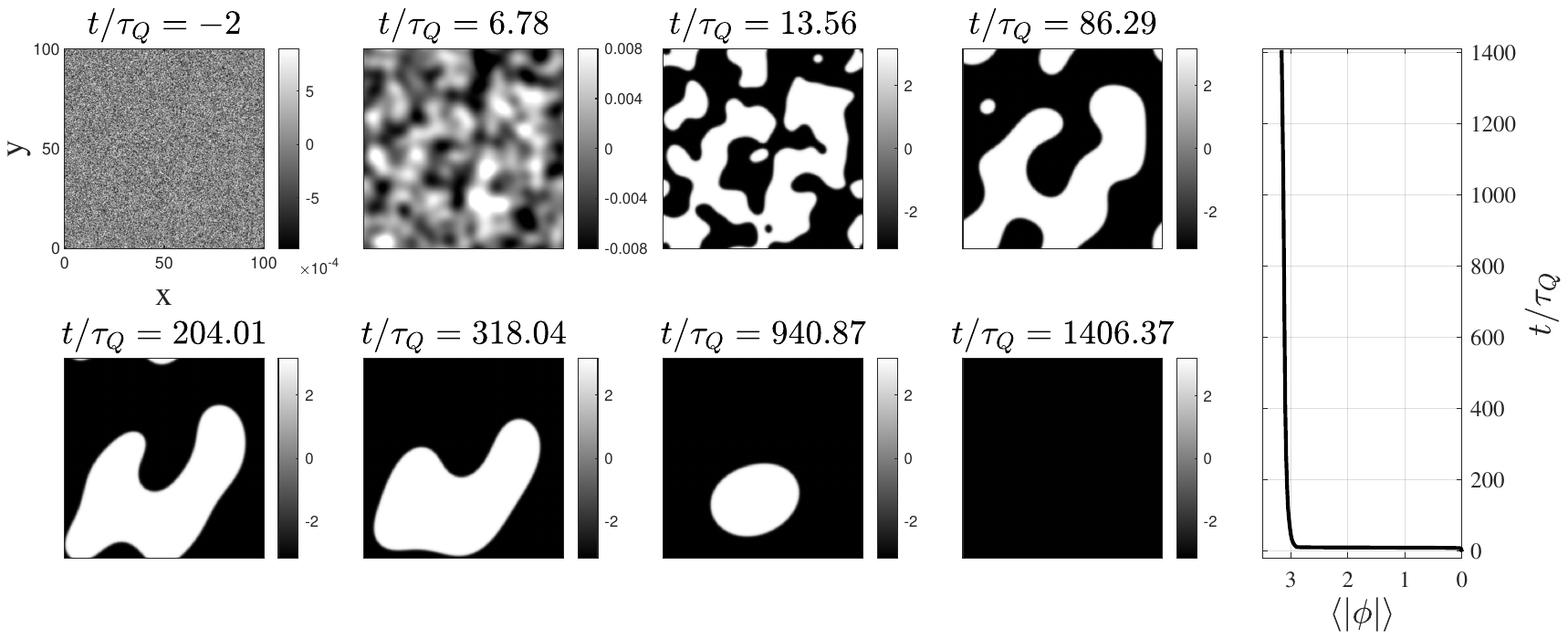}
\put(-420,162){(a)}
\put(-95,162){(b)}\\
\includegraphics[trim=4.8cm 14cm 4.4cm 13.7cm, clip=true, scale=1.2]{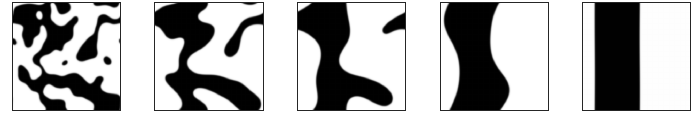}
\put(-410,70){(c)}\\
\includegraphics[trim=4.8cm 14cm 4.4cm 13.7cm, clip=true, scale=1.2]{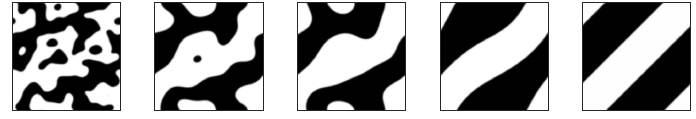}
\put(-410,70){(d)}
\caption{Snapshots of the density of order parameter at various time for periodic boundary conditions. (a) Time evolution of the domain walls without any topologically protected configurations. White and black domains represent the positive and negative order parameters, while the interfaces are the domain walls; (b) The corresponding time evolution of the average absolute values of the order parameter in panel (a); (c) Snapshots of the evolution of topologically protected domain walls with $(w_x,w_y)=(0,1)$; (d) Snapshots of the evolution of the topologically protected domain walls with $(w_x,w_y)=(1,1)$. }\label{w0}
\end{figure}

In Fig.\ref{w0}(a) we show various time snapshots of the densities of the order parameter, ranging from the initial state to the final equilibrium state. In the figure, white colors represent positive order parameter while black colors represent negative order parameters. Domain walls are the interfaces separating black and white areas. We can intuitively see how the order parameter evolves from random noise ($t/\tau_Q=-2$), to mosaic patterns ($t/\tau_Q=6.78$ and $t/\tau_Q=13.56$), and then the coarsening dynamics render the domain walls to bend and shrink (after $t/\tau_Q=86.29$). Finally domain walls disappear and leave the uniform negative order parameters in the space. 
We can see that the evolution of domain walls are in agreement with our speculations, from which the interfaces will bend and shrink like a rubber band with zero natural length. In particular, the uneven interface ($t/\tau_Q=204.01$) will gradually shrink and become much more even ($t/\tau_Q=940.87$), consistent with our illustrative plots in Fig.\ref{fig1}(c). Since there is no obstacle in Fig.\ref{w0}(a) to prevent the interface to shrink, it will finally disappear without any domain walls in the space. 
In our notations, the final configuration of the domain wall is $(w_x,w_y)=(0,0)$. The average free energy density of the final state is $f_{gs}=F[\phi]/V=-12.5$, where $V=L^2$ is the volume of the system. { (Please find the details of the free energy density in the Appendix A.) }  This configuration has the lowest free energy, thus, is the ground state and thermodynamically stable. 

Fig.\ref{w0}(b) exhibits the evolution of the average of the absolute values of the order parameter. It starts with very tiny order parameters for a while and then ramp quickly to a plateau. It should be noted that during the course of the plateau the system is still evolving although very slowly. The coarsening dynamics in this stage will gradually shrink the domain walls, thus the values of $\langle|\phi|\rangle$ will gradually grow. It will arrive at $\langle|\phi|\rangle\approx\sqrt{\epsilon_f}=\sqrt{10}$ until the final time of vanishing domain walls, 

There are possibilities that the domain walls may wind around the torus and thus it cannot disappear due to the topological obstacles, as Fig.\ref{w0}(c) and Fig.\ref{w0}(d) show. For convenience, we only show five snapshots of the time evolution (from left to right) of the order parameter. From Fig.\ref{w0}(c) we see that during the coarsening dynamics the domain walls will shrink, and topologically trivial domain walls will finally disappear just as in Fig.\ref{w0}(a). However, it is interesting to see that at the final stage, there exists nontrivial configuration of the domain wall with $(w_x,w_y)=(0,1)$, which is similar to the green loop in Fig.\ref{fig1}(b). \footnote{ One can read the winding numbers of a topologically protected metastable state directly. Taking the last snapshot in Fig.\ref{w0}(c) as an example. We can start at some point of the interface, then follow the route of the interface. Since this plot is periodic in $x$ and $y$ direction, finally we will return to the starting point. During this return, we wrap the $y$ direction once while wrap the $x$ direction zero times. Thus the winding number for this configuration is $(w_x,w_y)=(0,1)$. We will stress that we only count the absolute values of the windings, i.e., we do not distinguish the directions how we wrap around the spatial directions. Because direction is not important to our conclusions in the paper.} In Fig.\ref{w0}(c), the final state of the domain walls are straight lines, like rubber bands tightly winding around the $y$-direction, otherwise they will move according to curvatures.  From numerics we also find that this configurations of straight domain walls will exist permanently, meaning that they are dynamically stable. But from thermodynamics this configuration is an excited state rather than ground state. Since there is gradient energy existing in the domain walls. Specifically, the average free energy density of the final configuration is $f_{01}\approx -12.0773$ which is higher than that of the ground state $f_{gs}$. Therefore, from thermodynamics this final configuration is metastable. Thus, we have numerically demonstrated a topologically protected metastable state, which is in fact dynamically stable. 

Remarkably, it is still possible to have topologically protected metastable state having higher free energies. Fig.\ref{w0}(d) shows the final domain wall with configurations $(w_x,w_y)=(1,1)$, which winds along $x$ and $y$-directions once respectively. This configuration corresponds to the red curve in Fig.\ref{fig1}(b).  The system now is a higher excited state than that in Fig.\ref{w0}(c) since the length of interfaces is longer than those in Fig.\ref{w0}(c). Because of the periodic boundary conditions and the identical lengths of $x$ and $y$-directions, the interfaces need to have an angle of $\pi/4$ against the $x$-direction. This is exactly the case in the Fig.\ref{w0}(d). Numerically we get the average free energy density of the final state in this case as $f_{11}\approx-11.9037$, which is indeed greater than $f_{01}$ in Fig.\ref{w0}(c). 

Interestingly, we find that the difference of the average free energy density between the excited states in Fig.\ref{w0}(c) to the ground states in Fig.\ref{w0}(a) is $\Delta f_1=f_{01}-f_{gs}=0.4227$. On the other hand, the difference of $f$ between the higher excited state in Fig.\ref{w0}(d) and the ground state is $\Delta f_2=f_{11}-f_{gs}=0.5963$. The ratio $\Delta f_2/\Delta f_1\approx 1.4107\simeq\sqrt2$, which is the ratio between the length of the domain wall in Fig.\ref{w0}(d) and Fig.\ref{w0}(c). This in turn verifies that higher free energy configurations have longer domain walls, and the energy difference is linearly proportional to the lengths of the domain wall. \footnote{The codes for generating Fig.\ref{w0} can be downloaded from this \href{https://bhpan.buaa.edu.cn/link/AAED582A984A21410691F434D9429B0A0C}{link}. }

\subsection{Domain walls in two dimensions: II. Neumann Boundary Conditions}
%{\it Domain walls in two dimensions: II. Neumann Boundary Conditions} --  
At first sight, one may think that with Neumann boundary conditions there would not be topologically protected states, since the background geometry is trivial.  However, as we shall see there is still the chance to realize the topologically protected domain walls with Neumann boundary conditions. In numerics, we use the finite difference method to decompose the two spatial directions into $401$ grids, respectively.  

\begin{figure}[t]
\centering
\includegraphics[trim=0.2cm 3.5cm 0.cm 1.5cm, clip=true, scale=0.7]{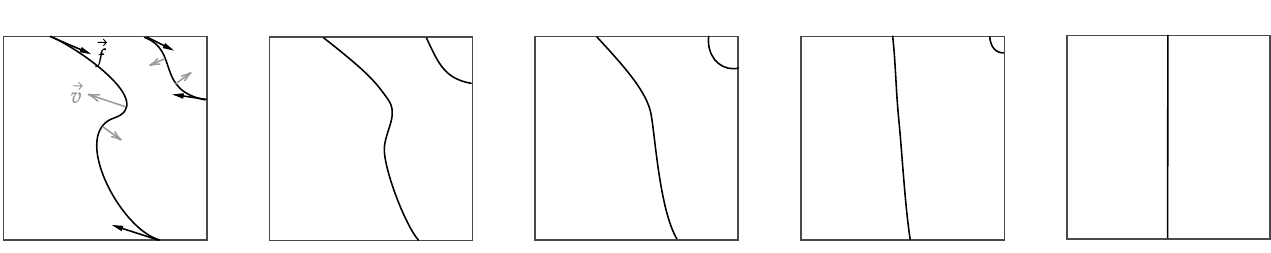}\put(-450,70){(a)}\\
\includegraphics[trim=3.7cm 13.5cm 2.7cm 13.5cm, clip=true, scale=1.045]{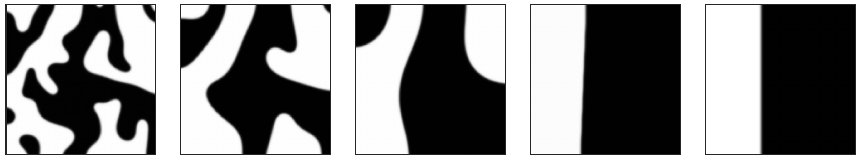}\put(-450,70){(b)}
\caption{Snapshots of the evolution of the order parameter with Neumann boundary conditions. (a) Sketchy picture of the evolution of the domain walls. Black arrows represent the force acting at the boundary points while the gray arrows indicate the velocity vectors. The domain wall connecting the opposite boundaries will finally become straight and stay for ever. However, the domain wall in the corner (connecting the adjacent boundaries) will shrink and finally disappear; (b) Numerical results demonstrating the topologically protected domain walls having configuration $(w_x,w_y)=(0,1)$ with Neumann boundary conditions. }\label{neumannpic}
\end{figure}

In Fig.\ref{neumannpic}(a) we illustrate why topologically nontrivial domain walls can exist with Neumann boundary conditions. The interface has surface tension, therefore, the force (black arrows) acting at the boundary points are pointing along the tangent direction of the interface. Due to the combination of the force at the boundary points and the velocity driven by the curvature, the interface which connects the opposite sides of the boundary will arrange itself and finally become a straight line perpendicular to the opposite boundaries.  It should be noted that with Neumann boundary conditions domain walls do not need to appear in pairs comparing to periodic boundary conditions. Therefore, we use $(w_x,w_y)$ to indicate the whole winding number of domain walls directly, rather than the half of it which was used in periodic boundary conditions. 

Fig.\ref{neumannpic}(b) shows a topologically nontrivial domain wall having configuration $(w_x, w_y)=(0,1)$ in the final equilibrium state with Neumann boundary conditions. Therefore, we demonstrate that even in the seemingly topologically trivial background geometry, we can still have the topologically protected metastable states. It is consistent with our speculation that the domain walls are like the rubber bands, since in this case the rubber bands connecting the two moving points at the opposite sides will try to drag them into a straight line perpendicular to the boundaries.  Consequently, the final configuration of the fields is symmetric in $y$-direction, which resembles a cylinder if we wrap it along $y$-direction. Therefore, the interface is like a rubber band winding along the azimuthal direction of the cylinder, which is topologically protected due the hole of the cylinder.  
The average free energy density of the final state is $f^N_{01}\approx -12.2906$, which is higher than that of the ground state in Fig.\ref{w0}(a). The difference between them is $\Delta f_0\approx f^N_{01}-f_{gs}=0.2094$, which is roughly the half of $\Delta f_1$, {\it viz.}, $\Delta f_1/2=0.21135$ since there is a pair of domain walls with periodic boundary conditions. 

However, the interfaces connecting the adjacent boundaries, for instance the curve in the corner in Fig.\ref{neumannpic}(a), will shrink and finally disappear due the joint effect of the force at the boundary points and the velocities from the curvature. They are not the topologically protected domain walls. Fig.\ref{neumannpic}(b) also shows the numerical evidences for this kind of domain walls.

\subsection{Domain walls in three dimensions}
\begin{figure}[t]
\centering
\includegraphics[trim=1.1cm 9.3cm 1.cm 8.7cm, clip=true, scale=0.8]{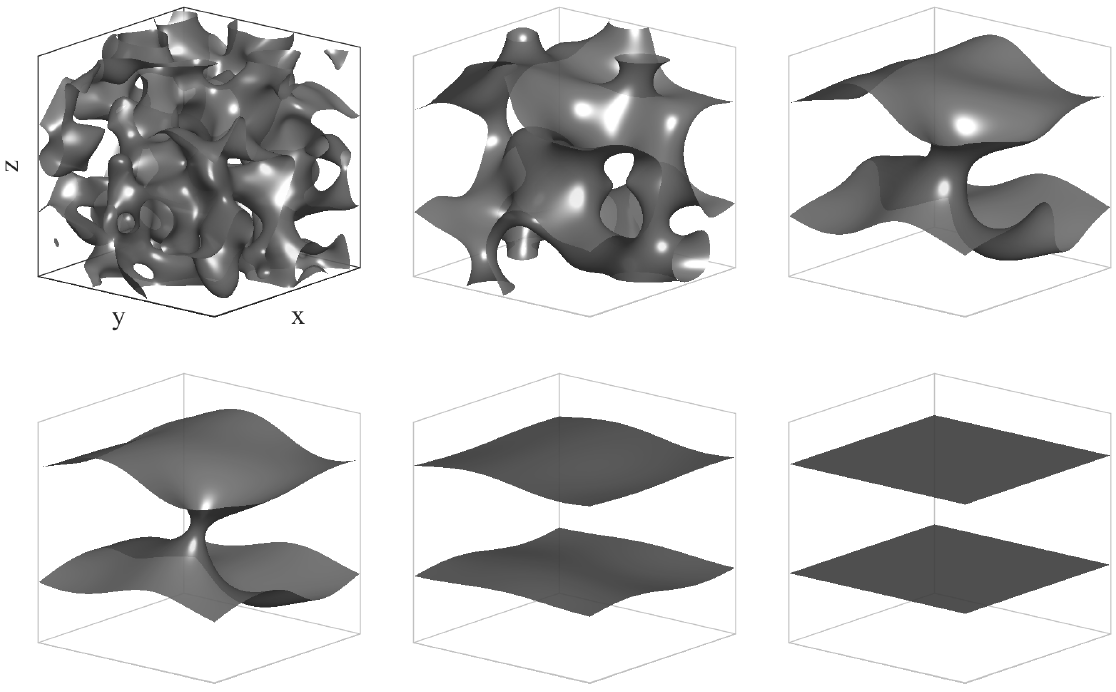}
\put(-420,260){(a)}\put(-275,260){(b)}\put(-130,260){(c)}
\put(-420,120){(d)}\put(-275,120){(e)}\put(-130,120){(f)}
\caption{Snapshots of the evolution ((a)$\to$(f)) of the isosurface of zero values of the order parameter in three dimensions with periodic boundary conditions.  }\label{3d}
\end{figure}
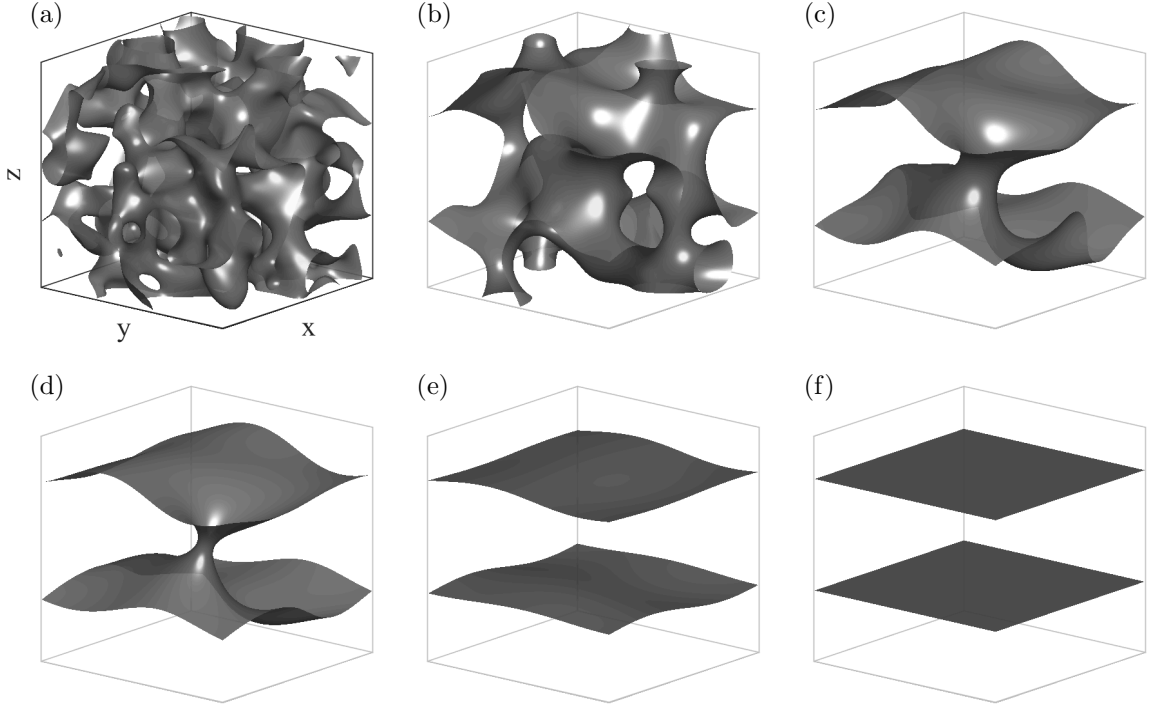

%{\it Domain walls in three dimensions} -- 
 In three dimensions it is difficult to show the density of order parameter in the whole bulk, alternatively, we can show the isosurface of zero values of the order parameter. Since the zero values separate the positive and negative order parameters, we can regard the zero value isosurface as a representative of the domain wall. In Fig.\ref{3d} we exhibit the snapshots of the evolution of the zero value isosurfaces in three dimensions with periodic boundary conditions. At the final time, there exists two topologically nontrivial flat surfaces indicating the topologically protected domain walls in three dimensions. One of the domain walls has the configuration $(w_x, w_y, w_z)=(1,1,0)$. Therefore, the topology of these domain walls are similar to those in Fig.\ref{w0}(c) although they are in different dimensions. The topology of domain walls similar to those in Fig.\ref{w0}(d) are hard to find in numerics. We believe that the probability is very low for this kind of skew surfaces. But Fig.\ref{3d} already gave a strong evidence for the existence of a topologically protected metastable state in three dimensions. 
The average free energy density now is $f_{110}\approx -11.6525$ which is higher than the ground state. The difference between them is $\Delta f_3=f_{110}-f_{gs}=0.8475$, which is roughly the double of $\Delta f_1$, {\it viz.}, $2\Delta f_1=2\times 0.4227=0.8454$. This is because now in three dimensions there is another pair of windings of the domain wall along $x$-direction compared to that in Fig.\ref{w0}(c). Therefore, we have realized the topologically protected metastable states in three dimensions. 

\section{Conclusions and Discussions}

%{\it Conclusions} -- 
We speculated that the structures of the domain walls in the TDGL model at the final equilibrium state can have nontrivial topologies even though they are metastable states from the aspects of thermodynamics. As far as we know, this was the first time to realize the explicit forms of the topologically protected metastable states in classical dynamics. Our speculations mainly relied on the Allen-Cahn's assertion that the velocity of domain wall was proportional to the mean curvature at that point. Therefore, the domain walls behaved as rubber bands which could bend, evolve and finally tightly wind around the background geometry in a nontrivial way. We numerically demonstrated our conjecture in two and three dimensions by using periodic or Neumann boundary conditions.   

{ It is well known that when a system is quenched across the critical point, the generated defects will satisfy the seminal Kibble-Zurek mechanism, i.e., the mean value of the defects number has a universal power-law to the quench rate \cite{kibble1976topology,kibble1980some,zurek1985cosmological}. In our paper, the defect number is the length of the domain wall since it is a one-dimensional object. However, as we see from the time evolution of the domain walls, they will experience coarsening dynamics at late time. Therefore, the lengths of the domain walls are changing in time! Finally, they will disappear if the configuration is trivial or they will have stable winding states if the configuration is topologically protected. This phenomenon is ubiquitous for other quench rates. Therefore, we expect that due to the coarsening dynamics the Kibble-Zurek mechanism will be destroyed at late time, which enriches the studies beyond the conventional Kibble-Zurek mechanism \cite{du2023kibble,del2022locality,gomez2022role}.}

In our model, we turned off the temperature of the system. However, temperature is not important to our conclusions. In fact we already checked that turning on the temperature by adding the Langevin noise term in the right side of Eq.\eqref{eqgl}, 
this model can still have the topologically protected metastable states, indicating that the existence of topologically protected metastable state has nothing to do with the temperature. As we have discussed, this metastable states protected by topology may potentially have practical use in the computer and information technology industry, such as the storage media since it has stable configurations of positive and negative order parameters.

\section*{Acknowledgements}

We thank Dr. Xin-Meng Wu for helpful discussions. This work was partially supported by the National Natural Science Foundation of China (Grants No. 12175008).

\appendix
\setcounter{equation}{0}
\setcounter{figure}{0}
\setcounter{table}{0}
\setcounter{section}{0}
%\setcounter{page}{1}
%\makeatletter
\renewcommand{\theequation}{S\arabic{equation}}
\renewcommand{\thefigure}{S\arabic{figure}}

{\section{Time evolution of free energy densities}}
\label{appa}

{
Free energy of the TDGL model can be found in Eq.\eqref{freeenergy} in the main text. In order to show that the final states in Fig.\ref{w0}(c), Fig.\ref{w0}(d), Fig.\ref{neumannpic}(b) and Fig.\ref{3d} are `dynamically stable' but `thermodynamically metastable', we plot the time evolution of the free energy densities $F/V$ of these states in the following Fig.\ref{freeE}. In this figure, the line $f_{gs}$ corresponds to the time evolution of free energy density of the ground state in Fig.\ref{w0}(a); $f_{01}$ corresponds to the states in Fig.\ref{w0}(c); $f_{11}$ corresponds to the states in Fig.\ref{w0}(d); $f_{01}^N$ corresponds to the states with Neumann boundary conditions in Fig.\ref{neumannpic}(b) and $f_{110}$ corresponds to the  states in Fig.\ref{3d}. In the horizontal axis of Fig.\ref{freeE} we use the dimensionless time coordinates $t/t_e$ where $t_e$ is the final time of the evolution for each state. Specifically, $t_e=(3826.6, 6902.6, 14403, 3600, 2496.6)$ for $(f_{gs}, f_{01}, f_{11}, f_{01}^{N}, f_{110})$, respectively. 

As shown in Fig.\ref{freeE}, the free energy densities will drop rapidly from zero in the early time, which is caused the the rapid quench. Then the free energy densities will experience slow declines and finally reach plateaus, indicating that finally they are dynamically stable. In the inset plot, we have enlarged the free energy densities in the late time in order to see the final free energy densities of each state clearly. As we already presented in the main text, the free energy densities of $ f_{01}, f_{11}, f_{01}^{N}$ and $f_{110}$ are all greater than that of the ground state $f_{gs}$, meaning that they are thermodynamically metastable. From this inset plot we can also vividly see the spacings between these free energy densities. For instance, there are equal spacings between $f_{01}$, $f_{01}^N$ and $f_{gs}$, indicating that $f_{01}$ has twice energy difference than that of $f_{01}^N$ to the ground state. As we already discussed in the main text, this twice difference corresponds exactly to the two times of the lengths of the domain walls of these final states. The rest free energy densities are already discussed in the main text as well, we will not repeat it here. 
  
%As can be seen from the figure, the free energy density first drops rapidly from zero, which is caused by rapid quenching. Then the free energy experienced a slow decline and finally reached a stable value, at which time the system underwent a long coarsening process. When the system is two-dimensional, we find that the equilibrium free energy density is $f_{11}$,$f_{01}$,$f_{01}^{N}$ and $f_{gs}$ in descending order, which corresponds exactly to the length of the domain wall of the final state.}

\begin{figure}[h]
	\centering
	\includegraphics[trim=1.1cm 9.3cm 1.cm 9.8cm, clip=true, scale=0.8]{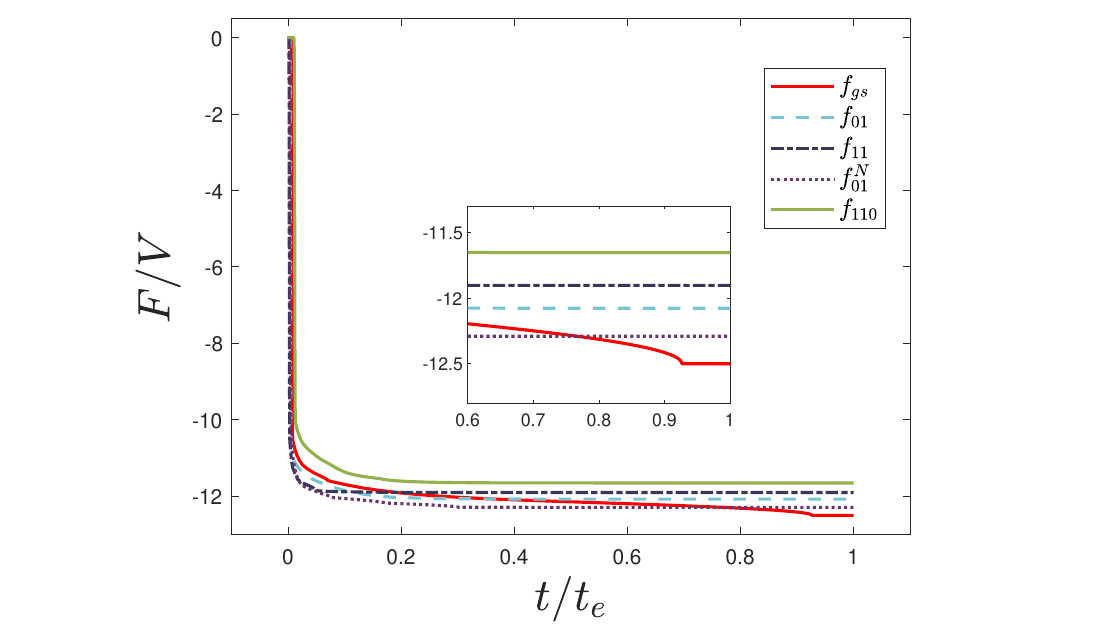}
	\caption{ Time evolution of free energy densities for the states in the main text. $t_e$ is the final time of evolution for each state. The inset exhibits the enlarged plot of the free energy densities in the late time. }\label{freeE}
\end{figure}

{\section{Topologically protected metastable states at finite temperature}}
\label{appb}

{When considering the temperature effect, we can add the Langevin noise term $\theta(t,\vec x)$ to the right side of the TDGL Eq.\eqref{GLeq2}, 
\be\label{}
\dot\phi-\Gamma\left(\nabla^2\phi-\frac12\left(\phi^3-\epsilon(t)\phi\right)\right)=\theta(t,\vec x), 
\ee
The noise term satisfies $\langle \theta(t,\vec x)\rangle=0$ and $\langle \theta(t',\vec x')\theta(t,\vec x)\rangle=2T\delta(\vec x'-\vec x)\delta(t'-t)$. In the numerics we set $T=0.001$ representing the temperature of the reservoir. Other numerical parameters are similar to those in the Fig.\ref{w0} in the main text. Fig.\ref{w1T} exhibits six snapshots of the order parameter at finite temperature.  We find that the system can still form a topologically protected metastable state with $(w_x,w_y)=(1,0)$ when the temperature is nonzero, meaning that this state is robust under temperature perturbations.}

\begin{figure}[h]
	\centering
	\includegraphics[trim=4.8cm 14cm 4.4cm 13.7cm, clip=true, scale=1.2]{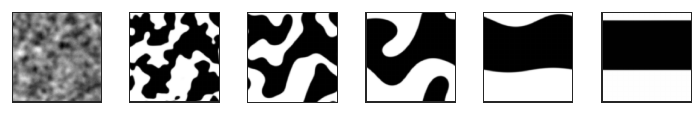}
	\caption{Snapshots of the evolution of topologically protected domain walls with $(w_x,w_y)=(1,0)$ at finite temperature. }\label{w1T}
\end{figure}
 }

\bibliographystyle{ieeetr} 
\bibliography{ref_domainwall.bib}

\begin{thebibliography}{10}

\bibitem{fefferman2017topologically}
C.~Fefferman, J.~Lee-Thorp, and M.~Weinstein, {\em Topologically protected
  states in one-dimensional systems}, vol.~247.
\newblock American Mathematical Society, 2017.

\bibitem{zeng2019quantum}
B.~Zeng, X.~Chen, D.-L. Zhou, and X.-G. Wen, {\em Quantum information meets
  quantum matter: From quantum entanglement to topological phases of many-body
  systems}.
\newblock Springer, 2019.

\bibitem{nayak2008non}
C.~Nayak, S.~H. Simon, A.~Stern, M.~Freedman, and S.~D. Sarma, ``Non-abelian
  anyons and topological quantum computation,'' {\em Reviews of Modern
  Physics}, vol.~80, no.~3, p.~1083, 2008.

\bibitem{stern2013topological}
A.~Stern and N.~H. Lindner, ``Topological quantum computation---from basic
  concepts to first experiments,'' {\em Science}, vol.~339, no.~6124,
  pp.~1179--1184, 2013.

\bibitem{hohenberg2015introduction}
P.~C. Hohenberg and A.~P. Krekhov, ``An introduction to the ginzburg--landau
  theory of phase transitions and nonequilibrium patterns,'' {\em Physics
  Reports}, vol.~572, pp.~1--42, 2015.

\bibitem{tang1995time}
Q.~Tang and S.~Wang, ``Time dependent ginzburg-landau equations of
  superconductivity,'' {\em Physica D: Nonlinear Phenomena}, vol.~88, no.~3-4,
  pp.~139--166, 1995.

\bibitem{Kopnin2001}
N.~B. Kopnin, ``{The Time-dependent Ginzburg-Landau Theory},'' in {\em {Theory
  of Nonequilibrium Superconductivity}}, Oxford University Press, 05 2001.

\bibitem{kibble1976topology}
T.~W. Kibble, ``Topology of cosmic domains and strings,'' {\em Journal of
  Physics A: Mathematical and General}, vol.~9, no.~8, p.~1387, 1976.

\bibitem{kibble1980some}
T.~W. Kibble, ``Some implications of a cosmological phase transition,'' {\em
  Physics Reports}, vol.~67, no.~1, pp.~183--199, 1980.

\bibitem{zurek1985cosmological}
W.~H. Zurek, ``Cosmological experiments in superfluid helium?,'' {\em Nature},
  vol.~317, no.~6037, pp.~505--508, 1985.

\bibitem{bray2002theory}
A.~J. Bray, ``Theory of phase-ordering kinetics,'' {\em Advances in Physics},
  vol.~51, no.~2, pp.~481--587, 2002.

\bibitem{hohenberg1977theory}
P.~C. Hohenberg and B.~I. Halperin, ``Theory of dynamic critical phenomena,''
  {\em Reviews of Modern Physics}, vol.~49, no.~3, p.~435, 1977.

\bibitem{allen1979microscopic}
S.~M. Allen and J.~W. Cahn, ``A microscopic theory for antiphase boundary
  motion and its application to antiphase domain coarsening,'' {\em Acta
  metallurgica}, vol.~27, no.~6, pp.~1085--1095, 1979.

\bibitem{vachaspati2007kinks}
T.~Vachaspati, {\em Kinks and domain walls: An introduction to classical and
  quantum solitons}.
\newblock Cambridge University Press, 2007.

\bibitem{weinberg1992classical}
E.~J. Weinberg, ``Classical solutions in quantum field theories,'' {\em Annual
  Review of Nuclear and Particle Science}, vol.~42, no.~1, pp.~177--210, 1992.

\bibitem{del2018universal}
A.~Del~Campo, ``Universal statistics of topological defects formed in a quantum
  phase transition,'' {\em Physical Review Letters}, vol.~121, no.~20,
  p.~200601, 2018.

\bibitem{PhysRevLett.124.240602}
F.~J. G\'omez-Ruiz, J.~J. Mayo, and A.~del Campo, ``Full counting statistics of
  topological defects after crossing a phase transition,'' {\em Phys. Rev.
  Lett.}, vol.~124, p.~240602, Jun 2020.

\bibitem{delCampo:2021rak}
A.~del Campo, F.~J. G\'omez-Ruiz, Z.-H. Li, C.-Y. Xia, H.-B. Zeng, and H.-Q.
  Zhang, ``{Universal statistics of vortices in a newborn holographic
  superconductor: beyond the Kibble-Zurek mechanism},'' {\em JHEP}, vol.~06,
  p.~061, 2021.

\bibitem{du2023kibble}
K.~Du, X.~Fang, C.~Won, C.~De, F.-T. Huang, W.~Xu, H.~You, F.~J.
  G{\'o}mez-Ruiz, A.~del Campo, and S.-W. Cheong, ``Kibble--zurek mechanism of
  ising domains,'' {\em Nature Physics}, pp.~1--7, 2023.

\bibitem{del2022locality}
A.~del Campo, F.~J. G{\'o}mez-Ruiz, and H.-Q. Zhang, ``Locality of spontaneous
  symmetry breaking and universal spacing distribution of topological defects
  formed across a phase transition,'' {\em Physical Review B}, vol.~106,
  no.~14, p.~L140101, 2022.

\bibitem{gomez2022role}
F.~J. G{\'o}mez-Ruiz, D.~Subires, and A.~del Campo, ``Role of boundary
  conditions in the full counting statistics of topological defects after
  crossing a continuous phase transition,'' {\em Physical Review B}, vol.~106,
  no.~13, p.~134302, 2022.

\end{thebibliography}

\end{document}